\documentclass{article} 
\usepackage{style/arxiv}

\usepackage{url}
\usepackage{import}
\usepackage{multirow}
\usepackage{subcaption}
\usepackage{graphicx}
\usepackage{amsmath}
\usepackage[table]{xcolor}
\usepackage{soul}
\usepackage{tabularx}
\usepackage{booktabs}
\usepackage{natbib}
\usepackage{xspace}
\usepackage{xcolor}
\usepackage{tabularx}

\usepackage[pagebackref=true,breaklinks=true,colorlinks,bookmarks=false]{hyperref}
\newcommand{\degs}{\textdegree/s}
\newcommand{\tablenote}[1]{\par\smallskip{\footnotesize\raggedright #1\par}}

\begin{document}

\title{From MNH to the Friedman--Komogortsev Method---which changes improve eye-movement classification?}
\renewcommand{\shorttitle}{The Friedman--Komogortsev Method for Eye-Movement Classification}

\author{
    Lee Friedman$^{\dagger}$, Mehedi Hasan Raju, Oleg Komogortsev\\
    Texas State University, San Marcos, Texas, USA\\
    {\small $^{\dagger}$Deceased}\\
    {\small\texttt{m.raju@txstate.edu, ok@txstate.edu}}
}
\date{}

\maketitle
\begin{abstract}

The Friedman--Komogortsev Method (FKM) is a rule-based classifier of fixations, saccades, and post-saccadic oscillations in 1000-Hz recordings, developed from the modified Nystr\"om--Holmqvist algorithm (MNH) and used in several recent studies. FKM changes MNH in many ways at once: it centres MNH's Savitzky--Golay filter, adds a position-based test for whether an apparent post-saccadic oscillation continues the saccade, adds rejection tests, and changes several thresholds. We asked which of these changes alter agreement with human classification. Both algorithms were scored against twenty 26-s GazeBase reading recordings hand-classified by three trained raters, and each change was reverted in turn. FKM agreed better than MNH as MNH was used: saccade Cohen's $\kappa$ rose on all 20 recordings (median $+0.034$), and saccade boundaries moved from 2--3~ms late to the raters' sample. One change produced this: centring the filter, which removes a three-sample delay. Reinstating the delay returned FKM's saccade $\kappa$ to MNH's value, and centring MNH's filter alone made MNH match FKM on saccades and exceed it on post-saccadic events. The position-based test changed how saccades with several velocity peaks were divided, splitting 8\% of rater-labelled ``wavy'' saccades into a saccade and a spurious oscillation against 39\% for delay-corrected MNH. FKM's higher post-saccadic velocity floor traded recall for precision. The remaining changes had no measurable effect on these recordings. FKM's improvement over MNH is therefore a filter-alignment correction that MNH users can apply directly, plus a segmentation rule for multi-peak saccades.
\end{abstract}

\keywords{eye movements, event detection, saccades, post-saccadic oscillations, fixations, reading, manual classification, Cohen's kappa}

\section{Introduction}\label{sec:intro}

Event detection is the first step in nearly every analysis of an eye-movement recording, and systematic errors at event boundaries propagate into every measure derived from them \cite{andersson2017,startsev2023}. The modified Nystr\"om--Holmqvist algorithm \citep[MNH;][]{friedman2018} adapted the adaptive velocity-threshold detector of \cite{nystrom2010} to EyeLink 1000 reading data by narrowing its smoothing window, fixing its thresholds, and moving its saccade onsets. Its successor, the Friedman--Komogortsev Method (FKM), was written by the first author between 2019 and 2022 and has been used in several studies since \cite{friedman2024fixation,melnyk2025,melnyk2026}, but it has not been described in a journal.

FKM was a response to two observations in the 2018 report. MNH, like the original algorithm, applies its Savitzky--Golay differentiator \cite{savitzky1964} causally, which displaces every smoothed signal by half the filter window, although ``classification of the position trace is typically what is desired'' \citep[p.~1375]{friedman2018}. And saccade timing remained MNH's most frequent error, leading the authors to conclude that ``the position trace signals need to be taken into consideration as well as the velocity trace'' \citep[p.~1394]{friedman2018}. FKM therefore centres the filter and adds a position-based test at the end of each saccade. It also adds rejection tests for candidate events and changes several thresholds.

When an algorithm changes many things at once, its overall agreement with human raters cannot say which change mattered. Here we evaluate FKM against MNH and revert each of FKM's changes in turn, to identify which changes improve agreement with human classification, which change how events are segmented without changing overall agreement, and which have no measurable effect. The full description of FKM, its code, and the details of its implementation are given in the Supplementary Material.

\section{Method}

\subsection{Recordings and reference labels}

We used the twenty GazeBase reading recordings \cite{griffith2021} hand-classified by \cite{friedman2023}: monocular EyeLink 1000 recordings at 1000~Hz, one per participant, of silent reading of two quatrains, with the first 26,000 samples classified. Raters assigned every sample to fixation, saccade, post-saccadic event (PSE), blink, artifact, or unclassifiable, and marked saccades whose velocity profile had more than one peak as ``wavy.'' The raters' PSE category includes single post-saccadic signal changes as well as oscillations \cite{friedman2023}; the algorithms' post-saccadic oscillation (PSO) class is scored against it. Three raters, graduate students trained over seven rounds against a written rule set \cite{friedman2021rules}, are the reference throughout; rater 2's labels survive for 16 recordings, giving 56 rater-by-recording studies. A fourth classifier, the first author, led that training and designed FKM; his labels are used only as a check. The same recordings were used during FKM's development, so this is a development-set evaluation.

\subsection{From MNH to FKM}

MNH smooths position and computes velocity and acceleration with a second-order, 7-sample Savitzky--Golay filter applied causally. Saccade candidates are runs of more than two samples above 55~\degs. A saccade starts at the local velocity minimum before the 45~\degs\ crossing, stepped forward by up to four samples, and ends at the minimum after it, stepped back three samples unless a PSO follows. A PSO starts after that minimum and ends when six consecutive samples fall below a per-recording fixation-noise threshold; its peak must lie between 20~\degs\ and the saccade's peak. Fixations are the remaining runs of at least 30~ms with a position range of at most 2\textdegree.

FKM keeps this structure, the 55~\degs\ peak threshold, and the fixation-noise estimate. Table~\ref{tab:changes} lists what it changes, grouped by what each change acts on. The slope test is the main new rule. When a velocity minimum inside a saccade would start a PSO, FKM regresses smoothed position on time on the dominant channel, over the candidate and over the end of the saccade; if the two slopes agree within a tolerance, the candidate is treated as part of the saccade rather than as an oscillation. A saccade with two velocity peaks therefore stays one saccade when its position is still moving at the saccade's rate. The full algorithm, its parameters and output codes, and the corrections made to its code before this evaluation are in Supplementary Material S1--S3.

\begin{table}[!htbp]
\caption{What FKM changes in MNH}\label{tab:changes}
\small
\begin{tabular}{@{}>{\raggedright\arraybackslash}p{0.19\textwidth}
                  >{\raggedright\arraybackslash}p{0.27\textwidth}
                  >{\raggedright\arraybackslash}p{\dimexpr0.54\textwidth-4\tabcolsep\relax}@{}}
\toprule
Acts on & MNH & FKM \\
\midrule
Signal alignment & Savitzky--Golay filter (2nd order, 7 samples) applied causally: 3-sample delay & Same filter applied centred: no delay \\
Saccade boundaries & On/off threshold 45~\degs; onset up to 4 samples after the minimum & On/off threshold 40~\degs; onset 3 samples after the minimum \\
Saccade vs.\ PSO & Every interior velocity minimum can end a saccade; PSO rejected if its peak exceeds the saccade's & \textit{Slope test}: candidate PSO that continues the saccade's position trend is kept in the saccade \\
PSO acceptance & Peak $\geq$ 20~\degs; duration $\geq$ 3~ms & Peak $>$ 40~\degs; duration $\geq$ 7~ms; at most six velocity maxima; peaks must decline; at most 17~ms longer than the saccade \\
Saccade rejection & Missing data; duration $<$ 10~ms & Also: blink without missing samples; peak block $>$ 105~ms; another saccade in the preceding 20~ms; noisy baseline; velocity peak at first or last sample; duration $<$ 8~ms \\
Fixations & $\geq$ 30~ms; range $\leq$ 2\textdegree, else discarded & $\geq$ 20~ms; range $\leq$ 2.5\textdegree, else trimmed to its stable part \\
Noise & Velocity $>$ 1000~\degs; RIONEPS detector & Velocity $>$ 680~\degs; no RIONEPS detector \\
\bottomrule
\end{tabular}
\end{table}

\subsection{Evaluation}

\textit{Agreement.} We followed \cite{friedman2023}. With the rater as reference, scoring was restricted to runs of at least 300~ms in which both rater and algorithm assigned fixation, saccade, or PSO. Sample-level agreement was Cohen's $\kappa$ for each class; event-level agreement was F1, a rater event counting as a hit when its best intersection-over-union match exceeded 0.5 \cite{zemblys2018}. PSO metrics require at least 20 rater events, leaving 42 studies on 16 recordings. Human--human agreement was computed in the same way for every ordered pair of trained raters. Boundary displacement is the algorithm's onset or offset minus the rater's, in milliseconds, for matched events. A saccade is ``split'' when the algorithm labels part of the rater's saccade interval as PSO.

\textit{Ablations.} FKM was re-run with one change reverted to MNH's setting at a time: the filter applied causally (delay reinstated), the slope test disabled, the PSO floor lowered to 20~\degs, the onset step set to four samples, the on/off threshold raised to 45~\degs, and the minimum PSO duration lowered to 3~ms; a seventh run reverted all six. The rejection tests were not reverted individually; we report how often each fired.

\textit{Delay-corrected MNH.} To ask what FKM adds once MNH's delay is removed, we re-ran MNH from its source code with only the filter centred (Supplementary Material S4).

\textit{Statistics.} Paired comparisons used the samples both algorithms classified. The unit of inference is the recording: each recording's difference is the median over raters, and the 20 differences (16 for PSO) were tested with the exact two-sided sign test, with a percentile bootstrap 95\% confidence interval (10,000 resamples). No correction for multiple comparisons was applied, and the analyses were not preregistered.

\section{Results}

\subsection{FKM against MNH}

FKM's saccade $\kappa$ exceeded MNH's on all 20 recordings (median difference $+0.034$, 95\% CI $[+0.025, +0.044]$, $p < .001$). Fixation $\kappa$ ($+0.007$, $p = .82$) and saccade F1 did not differ clearly. PSO F1 was higher for FKM by a median of 0.081 but on only 10 of 16 recordings ($p = .45$): FKM produced 85\% fewer PSO false alarms but found fewer of the raters' PSEs (pooled precision 0.71 vs.\ 0.30; recall 0.38 vs.\ 0.43). MNH placed boundaries late by a median of 2~ms at saccade onset and offset and 3~ms at PSO onset; FKM placed them at $-1$, 0, and 0~ms (Table~\ref{tab:ablation}). The displacement was the same against each of the four raters, including the designer.

\subsection{The effect of each change}

\begin{figure}[!htbp]
\centering
\includegraphics[width=\textwidth]{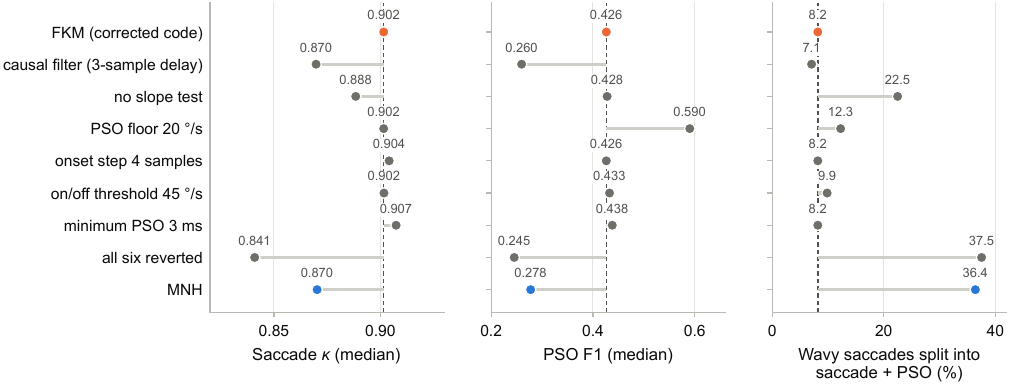}
\caption{FKM with each change reverted to MNH's setting (Table~\ref{tab:ablation}). The dashed line marks FKM. Left to right: median saccade $\kappa$, median PSO F1, and the percentage of rater-labelled wavy saccades split into a saccade and a PSO.}\label{fig:ablation}
\end{figure}

Table~\ref{tab:ablation} and Fig.~\ref{fig:ablation} show FKM with each change reverted. The changes separate into three kinds.

\begin{figure}[!htbp]
\centering
\includegraphics[width=\textwidth]{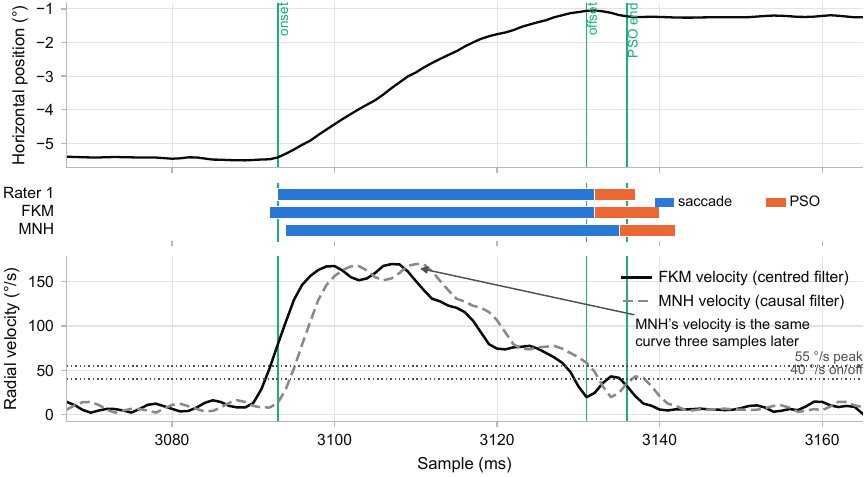}
\caption{A saccade and its post-saccadic event from recording S\_1051\_S1. Top: raw horizontal position. Middle: classification by rater~1, FKM, and MNH (blue, saccade; orange, PSO/PSE). Bottom: radial velocity from FKM's centred filter and MNH's causal filter; the curves are identical in shape and three samples apart. Vertical lines mark the rater's onset, offset, and PSE end.}\label{fig:delay}
\end{figure}

\textit{The change that improves agreement: signal alignment.} Reinstating the three-sample delay, and nothing else, lowered FKM's saccade $\kappa$ from 0.902 to 0.870, MNH's value (lower on 19 of 20 recordings, $p < .001$), moved its boundaries to $+2$, $+3$, and $+3$~ms, and lowered PSO F1 by a recording-level median of 0.087 (0 of 13 recordings higher, $p < .001$) and PSO precision from 0.72 to 0.55. Figure~\ref{fig:delay} shows why. MNH's velocity is FKM's velocity three samples later; because its smoothed position is delayed by the same amount, MNH's signals agree with one another and show nothing wrong, and the lag appears only against the raw recording.

\textit{Changes that alter segmentation: the slope test and the PSO floor.} Disabling the slope test nearly tripled the share of wavy saccades split into a saccade and a PSO (8.2\% to 22.5\%) and lowered PSO precision from 0.72 to 0.63, while changing saccade $\kappa$ by only $-0.005$ (5 of 19 recordings higher, $p = .06$) and PSO F1 by $-0.005$. Lowering the PSO floor from 40 to MNH's 20~\degs\ raised PSO $\kappa$ by 0.173 and PSO F1 by 0.208 (16 of 16 and 15 of 16 recordings), raising recall from 0.40 to 0.61 and lowering precision from 0.72 to 0.61. The 40~\degs\ floor is why FKM finds only 40\% of the raters' PSEs: it removed 963 of the 1811 candidates that passed FKM's other PSO tests.

\textit{Changes with little or no effect.} A four-sample onset step moved the median saccade onset from $-1$ to 0~ms with no change in agreement; raising the on/off threshold to 45~\degs\ changed no median by more than 0.007; lowering the minimum PSO duration to 3~ms raised fixation and saccade $\kappa$ slightly but consistently ($+0.006$, 20 of 20; $+0.004$, 18 of 20). Of FKM's new saccade rejection tests, two never fired on these recordings---no candidate exceeded 105~ms, and every blink produced missing samples---and the refractory, noisy-baseline, and peak-position tests together rejected 111 of 5704 candidates. The new PSO tests rejected 30 of 2134 logged candidates, and the 17-ms duration limit removed 14 more.

Reverting all six changes together produced MNH's PSO behaviour (precision 0.31, 37.5\% of wavy saccades split) but a lower saccade $\kappa$ than MNH (0.841 vs.\ 0.870). Part of that residual is MNH's variable onset step, which places onsets about one sample earlier than the fixed four-sample step used in this variant ($+2$ vs.\ $+3$~ms; Supplementary Material S4).

\begin{table}[!htbp]
\centering
\caption{FKM with each change reverted to MNH's setting}\label{tab:ablation}
\small
\setlength\tabcolsep{3.5pt}
\begin{tabular}{@{}lccccccc@{}}
\toprule
 & Fix. & Sac. & PSO & PSO & PSO prec. & Timing & Wavy \\
Variant & $\kappa$ & $\kappa$ & $\kappa$ & F1 & / recall & (ms) & split (\%) \\
\midrule
FKM & 0.867 & 0.902 & 0.482 & 0.426 & 0.72 / 0.40 & $-1$, 0, 0 & 8.2 \\
\ \ delay reinstated & 0.857 & 0.870 & 0.384 & 0.260 & 0.55 / 0.30 & $+2$, $+3$, $+3$ & 7.1 \\
\ \ slope test off & 0.863 & 0.888 & 0.476 & 0.428 & 0.63 / 0.40 & $-1$, 0, 0 & 22.5 \\
\ \ PSO floor 20~\degs & 0.876 & 0.902 & 0.635 & 0.590 & 0.61 / 0.61 & $-1$, 0, 0 & 12.3 \\
\ \ onset step 4 & 0.868 & 0.904 & 0.482 & 0.426 & 0.72 / 0.39 & 0, 0, 0 & 8.2 \\
\ \ on/off 45~\degs & 0.867 & 0.902 & 0.489 & 0.433 & 0.71 / 0.41 & $-1$, 0, 0 & 9.9 \\
\ \ min.\ PSO 3~ms & 0.875 & 0.907 & 0.486 & 0.438 & 0.72 / 0.40 & $-1$, 0, 0 & 8.2 \\
\ \ all six reverted & 0.840 & 0.841 & 0.473 & 0.245 & 0.31 / 0.42 & $+3$, $+3$, $+3$ & 37.5 \\
MNH as used & 0.862 & 0.870 & 0.476 & 0.278 & 0.30 / 0.43 & $+2$, $+2$, $+3$ & 36.4 \\
MNH, delay-corr. & 0.892 & 0.895 & 0.610 & 0.569 & 0.49 / 0.70 & $-1$, 0, 0 & 38.8 \\
\midrule
Trained rater pairs & 0.946 & 0.963 & 0.803 & 0.783 & & & \\
\bottomrule
\end{tabular}
\tablenote{Medians over 56 rater-by-recording studies (42 for PSO), each variant on the samples it classified, trained raters 1--3 as reference. Precision and recall are pooled. Timing: median algorithm minus rater boundary at saccade onset, saccade offset, and PSO onset. Wavy split: share of 538 rater-labelled wavy saccades split into saccade and PSO. MNH, delay-corr.: MNH re-run with its filter centred. Trained rater pairs: every ordered pair of raters 1--3.}
\end{table}

\subsection{What FKM adds once the delay is removed}

Centring MNH's filter, with no other change, raised MNH's saccade $\kappa$ by a median of 0.029 (19 of 20 recordings), fixation $\kappa$ by 0.029 (20 of 20), and PSO $\kappa$ and F1 by 0.127 and 0.247 (16 of 16), and placed its boundaries on the raters' sample (Table~\ref{tab:ablation}). Against this delay-corrected MNH, FKM showed no saccade advantage ($\kappa$ $+0.000$, 95\% CI $[-0.007, +0.013]$, 10 of 20 recordings), lower fixation $\kappa$ ($-0.026$, 2 of 20, $p < .001$), and lower PSO agreement ($\kappa$ $-0.139$, F1 $-0.152$; 2 of 16, $p = .004$). With MNH's 20~\degs\ floor, FKM and delay-corrected MNH did not differ clearly on PSO $\kappa$ ($-0.011$, $p = .21$) or F1 ($-0.002$, $p = .80$); two small differences reached $p < .05$ (fixation $\kappa$ $-0.008$, saccade F1 $+0.006$).

\begin{figure}[!htbp]
\centering
\includegraphics[width=\textwidth]{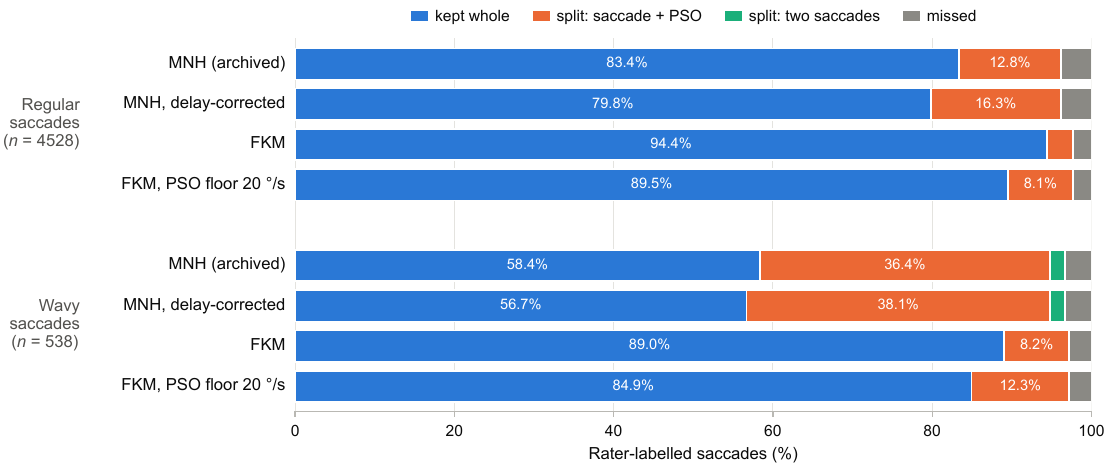}
\caption{Treatment of rater-labelled saccades (rater-event instances, trained raters pooled): kept whole, split into a saccade and a PSO, split into two saccades, or missed. ``Delay-corrected'' MNH here is the archived classification moved three samples earlier, which matches the centred re-run on 99.77\% of samples (38.1\% vs.\ 38.8\% of wavy saccades split; Supplementary Material S4).}\label{fig:wavy}
\end{figure}

What remained was segmentation (Fig.~\ref{fig:wavy}). Delay-corrected MNH split 38.8\% of wavy and 16.3\% of regular rater saccades into a saccade and a PSO; FKM split 8.2\% and 3.2\%, and 12.3\% and 8.1\% with the 20~\degs\ floor. A trained rater's PSE fell inside another trained rater's saccade in 3.3\% of wavy and 5.0\% of regular instances.

\section{Discussion}

Of the many changes FKM makes to MNH, one improves agreement with human classification: applying the Savitzky--Golay filter centred instead of causally. It accounts for FKM's whole saccade advantage over MNH as MNH was used, and it can be applied to MNH itself: centred, MNH agrees with the raters as well as FKM on saccades and better on fixations and post-saccadic events. A second change, the slope test, does not raise overall agreement but changes how saccades with several velocity peaks are divided, bringing FKM's treatment of them close to the raters'. A third, the higher PSO velocity floor, makes FKM's PSOs more precise and less complete. The remaining threshold changes and the new rejection tests had little or no measurable effect on these recordings.

The filter-alignment result applies beyond these two algorithms. A Savitzky--Golay filter is symmetric; applied causally, for example with MATLAB's \texttt{filter} or SciPy's \texttt{lfilter}, it delays every derived signal by $(F-1)/2$ samples for a window of $F$ samples, nine at the 19-sample window of \cite{nystrom2010} at 1000~Hz. Because position, velocity, and acceleration are delayed together, the detector's own signals remain mutually consistent, and durations, amplitudes, and peak velocities are unaffected; what changes is when each event occurs relative to the recording, the stimulus, and any other signal. Here a 2--3~ms lag cost 0.03 in saccade $\kappa$ and 0.13 in PSO $\kappa$. Users of velocity-based detectors should check how the differentiator is applied and, if causally, either subtract the group delay from every boundary or apply the filter centred; for MNH the two give the same result (Supplementary Material S4).

The slope test addresses a failure that sample-level agreement barely registers. A velocity rule ends a saccade at any interior velocity minimum, so a saccade with two velocity peaks is shortened, its amplitude underestimated, and its second half labelled a PSO, distorting main-sequence and PSO measures. FKM's PSO floor should be chosen for the purpose: 40~\degs\ when a detected PSO must be real, 20~\degs\ when agreement with PSE labels like these matters more, although the 20~\degs\ advantage was observed on the recordings used to develop FKM.

Several limitations apply. The evaluation uses FKM's development recordings, labelled by raters trained under rules its designer wrote, including the definition of a wavy saccade; agreement with them is agreement with one explicit convention \cite{hooge2018,hessels2018}. The data come from one tracker, one task, and young adults with good calibration, and two new rejection tests were never exercised, so their value on noisier data is untested. Like MNH, FKM uses a fixed 55~\degs\ peak threshold and so misses very small saccades \cite{friedman2018}. No rule-based detector approached human--human agreement on post-saccadic events, whose definition here includes non-oscillatory signal changes and whose form in video-based recordings partly reflects pupil motion \cite{nystrom2013,hooge2016}.

\bibliographystyle{unsrt}
\bibliography{ms.bib}

\end{document}


\title{Supplementary Material for \textit{From MNH to the Friedman--Komogortsev Method---which changes improve eye-movement classification?}}
\renewcommand{\shorttitle}{Supplementary Material for the Friedman--Komogortsev Method for Eye-Movement Classification}

\author{
    Lee Friedman$^{\dagger}$, Mehedi Hasan Raju, Oleg Komogortsev\\
    Texas State University, San Marcos, Texas, USA\\
    {\small $^{\dagger}$Deceased}\\
    {\small\texttt{m.raju@txstate.edu, ok@txstate.edu}}
}
\date{}

\maketitle
\thispagestyle{empty}

\section{Full description of FKM}\label{sec:fkm}

Lee Friedman designed and wrote FKM between 2019 and 2022 and died before this description was completed. The account below was reconstructed from his MATLAB source, his user guide, and the code comments, and every statement was verified against the code. Where a design choice is not explained in his materials, we say so rather than supply a rationale he did not record. ``P\textit{nn}'' refers to the numbered MATLAB functions of the distribution (P01--P24). All parameters are set for 1000-Hz data; several are expressed in samples and must be re-set for other sampling rates.

FKM keeps MNH's structure---noise first, then saccades and their PSOs, then fixations as the residue---and its central parameters (Table~\ref{tab:params}). Figure~\ref{fig:pipeline} gives the order of operations.

\begin{figure}[!htbp]
\centering
\small
\fbox{\begin{minipage}{0.94\textwidth}
\begin{enumerate}
\item \textbf{Signal conditioning (P06).} Centred Savitzky--Golay smoothing and differentiation; tracker NaN (subtype 5) and filter NaN (6) labelled noise.
\item \textbf{Fixation-noise threshold (P08).} 90th percentile of velocity in provisional fixations; the only per-recording adaptive quantity.
\item \textbf{Noise and artifact (P10).} Exclusion windows around NaN runs (9, 10); velocity above 680~\degs\ (7) or acceleration above 225,000~\degss\ (8).
\item \textbf{Saccades and PSOs (P11--P16),} one velocity peak block ($>55$~\degs) at a time:
  \begin{enumerate}
  \item peak-block tests: NaN, brevity, blink without NaN, duration, overlap with an accepted saccade or PSO;
  \item onset: back to 40~\degs, back to the local minimum, forward 3 samples; refractory and noisy-baseline tests;
  \item PSO candidate after the offset minimum: noise test, slope test pass 1, minimum duration, slope test pass 2, local-maxima and declining-peak tests;
  \item saccade offset fixed (by the PSO, the slope test, or the minimum $-$ 3 samples); whole-saccade tests: blink, already PSO, duration, interior peak, NaN;
  \item PSO kept only if its duration exceeds the saccade's by at most 17~ms and its peak exceeds 40~\degs.
  \end{enumerate}
\item \textbf{Fixations (P19).} Residual runs of at least 20~ms; over-range runs salvaged to their longest stable portion.
\item \textbf{Clean-up and checks (P20, P21, P24).} PSOs not preceded by a saccade removed; consistency check.
\end{enumerate}
\end{minipage}}
\caption{Order of operations in FKM. Numbers in parentheses are subtype codes (Table~\ref{tab:subtypes}). Labels are written once, so the order fixes the precedence among stages.}\label{fig:pipeline}
\end{figure}

\subsection{Signal conditioning}

FKM computes smoothed position, velocity, and acceleration in one pass from the Savitzky--Golay coefficient matrix. The signal is padded at each end with $(F-1)/2$ replicates of its first and last values, where $F = 7$ is the window length; each derivative column of the coefficient matrix, scaled by $p!/(-\Delta t)^{p}$ for derivative order $p$, is convolved with the padded signal; and the result is truncated to the original length (P06). This has zero group delay: the estimates at sample $n$ come from the window centred on $n$. Radial velocity and acceleration are the Euclidean norms of the horizontal and vertical components. Samples that are NaN in the raw signal are labelled as tracker loss (subtype 5); samples that become NaN through the filter---the window's reach around each tracker loss---are labelled separately (subtype 6).

\subsection{The fixation-noise threshold}

\begin{figure}[!htbp]
\centering
\includegraphics[width=0.9\textwidth]{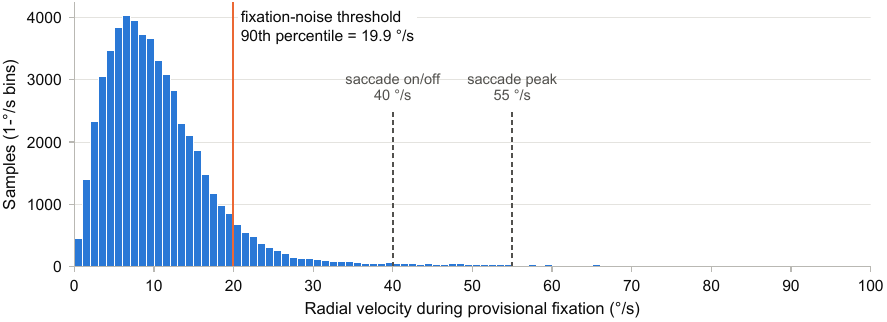}
\caption{Radial velocity during provisional fixations in recording S\_1051\_S1 (samples below 100~\degs, in runs of at least 20~ms, trimmed by 3 samples at each end). The 90th percentile (19.9~\degs) is FKM's fixation-noise threshold for this recording. The fixed saccade thresholds are shown for comparison.}\label{fig:fixnoise}
\end{figure}

FKM retains MNH's per-recording estimate of velocity noise during fixation, with the minimum run length reduced to 20~ms and the trim to 3 samples at each end (P08). The distribution is skewed (Fig.~\ref{fig:fixnoise}), and its 90th percentile---between 8.5 and 25.3~\degs\ across the twenty recordings---is the one adaptive quantity in FKM. It enters five decisions: the extension of the exclusion window around each tracker loss, the test of a candidate saccade's baseline, the end of a PSO, the extension of a saccade by the slope test, and the salvage of fixations.

\subsection{Noise and artifact}

The exclusion window around each run of tracker loss is extended exactly as in MNH---outward in 10-sample steps until a step's maximum velocity falls below the fixation-noise threshold---with pre- and post-blink samples labelled separately (subtypes 9 and 10). Within ten samples of the extended window, any sample whose acceleration exceeds the ceiling causes a further 21-sample window to be excluded, and the velocity and acceleration within it are clamped so that it cannot seed a later detection (P10). Runs of velocity above 680~\degs\ or acceleration above 225,000~\degss\ anywhere in the recording are labelled artifact (subtypes 7 and 8). The velocity ceiling is below the peak velocity of a large saccade; it was set for reading and must be raised for tasks with large saccades. FKM does not implement MNH's RIONEPS detector (Section~\ref{sec:rioneps}).

\subsection{Saccade detection}

Saccade candidates are runs of more than two consecutive samples above 55~\degs\ (``peak blocks''; P11). Each candidate passes through thirteen tests in sequence (P12); six, marked here with an asterisk, have no counterpart in MNH. The peak block is rejected if it contains a NaN; if it is too brief; if it is a blink that produced no missing samples*; if its last sample lies 105 or more samples after its first*; or if it overlaps an accepted saccade or an accepted PSO. The blink test is geometric: in a block longer than 25 samples, the minimum of the vertical position is located; if it falls between 30\% and 70\% of the way through the block, and the rises in vertical position before and after it both exceed 4\textdegree\ with a ratio below 2, the block is a blink (subtype 16). The test is applied again to the assembled saccade.

Onset and offset are found as in MNH, with an on/off threshold of 40~\degs\ and a forward step of three samples; the backward step of three samples at offset is applied only when no PSO follows. The PSO detector runs before the offset is fixed. The assembled saccade is rejected if any of the 20 samples before its onset is already labelled saccade*; if more than 15 of those 20 samples exceed the fixation-noise threshold and none is a PSO* (subtype 19);\footnote{The code divides the count of noisy samples by 30, a remnant of MNH's 30-ms minimum fixation, and compares the result with 0.5; with a 20-sample window this requires more than 15 noisy samples.} if it is already labelled PSO*; if it is shorter than 8~ms; if its maximum velocity falls on its first or last sample* (subtype 18); or if it contains a NaN. Accepted saccades with amplitude above 8.08\textdegree\ are labelled large (subtype 4), a criterion chosen to separate return sweeps from within-line saccades in reading. The rationale for the 105-ms maximum and the 8.08\textdegree\ criterion is not recorded beyond their being chosen on GazeBase reading data.

\subsection{PSO detection and the slope test}

A PSO candidate begins at the sample after the saccade's local velocity minimum. Its end is located from the first sample $m$ at which six consecutive samples ($m$ to $m+5$) all fall below the fixation-noise threshold; the end index is $m+1$ (P13, P14). It is rejected if any sample in it is already labelled noise.

\begin{figure}[!htbp]
\centering
\includegraphics[width=\textwidth]{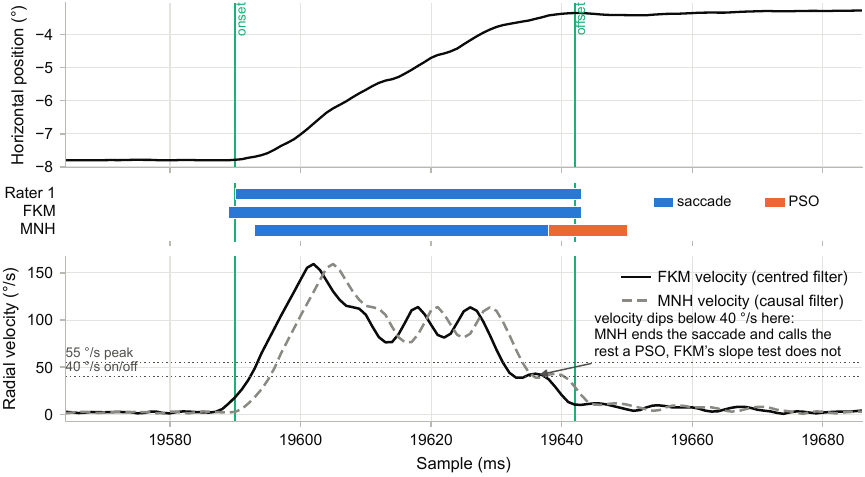}
\caption{A saccade that rater 1 labelled ``wavy,'' from recording S\_1092\_S2. Radial velocity (bottom) has three peaks and dips below the 40~\degs\ on/off threshold at sample 19636, where MNH ends the saccade and labels the remaining travel a PSO (bars, top). The position trace is still moving at that point at a rate indistinguishable from the saccade's, and FKM's slope test keeps the samples in the saccade, as the rater did.}\label{fig:wavyex}
\end{figure}

The slope test (P16) addresses saccades whose velocity profile has two peaks, whose interior velocity minimum any velocity-based rule takes for the end of the saccade. MNH rejected any candidate whose peak velocity exceeded that of its saccade, which discards the samples along with the error. FKM examines position instead. For the saccade and for the candidate independently, the channel (horizontal or vertical) with the larger displacement is identified; if they differ, the test does not apply. If they agree, smoothed position on that channel is regressed on sample index over the candidate and over the terminal segment of the saccade of the same length (\texttt{regress}, $\alpha = 0.0001$), in two passes:
\begin{itemize}
\item \textit{Pass 1}, only when the candidate's peak velocity exceeds the saccade's: the candidate continues the saccade if the slopes have the same sign, both exceed 0.02~\textdegree/sample (20~\degs) in magnitude, and the candidate's slope lies within the saccade's confidence interval widened by 0.05~\textdegree/sample on each side.
\item \textit{Pass 2}, for every candidate that survives pass 1 and is at least 7~ms long: the candidate continues the saccade if the saccade's slope has the same sign as the candidate's, the candidate's slope exceeds 0.02~\textdegree/sample in magnitude, and it lies within the saccade's confidence interval widened by 0.03~\textdegree/sample on each side; the slopes are then recomputed over the extended candidate, and the extension is made only if the dominant channel is unchanged and the extended candidate is more than four samples long.
\end{itemize}
When the test is met, the saccade's end is moved to the first sample at which velocity falls to the fixation-noise threshold, snapped to the nearest position extremum within seven samples, and no PSO is recorded. The confidence interval is used as a tolerance, not as an inferential test: consecutive samples of smoothed position are strongly autocorrelated. Figure~\ref{fig:wavyex} shows the test on a rater-labelled wavy saccade.

Candidates shorter than 7~ms are discarded between the two passes. Two morphological tests follow (P15): a candidate with more than six local velocity maxima is rejected as oscillatory noise; and a candidate whose first velocity peak exceeds 40~\degs, occurs more than four samples after onset, and is exceeded by a second peak more than five samples later by a factor above 1.7 is rejected as not a damped oscillation (subtype 17). Once the saccade is accepted, the PSO is kept only if it does not exceed the saccade's duration by more than 17~ms and its peak velocity exceeds 40~\degs\ (P11). Accepted PSOs are graded small, moderate, or large at 40, 51, and 72~\degs. After detection, a sweep removes any PSO whose preceding sample is not a saccade (P21).

\subsection{Fixation detection}

Fixations are the residue: runs of samples labelled neither saccade, PSO, nor noise (P19). A run shorter than 20~ms is rejected (subtype 14). A run whose position range---the larger of the horizontal and vertical ranges---exceeds 2.5\textdegree\ is salvaged rather than discarded. Samples above the fixation-noise threshold are marked as high-velocity, runs of fewer than five such samples are unmarked so that brief noise does not fragment the fixation, and samples above 55~\degs\ are marked additionally, with runs of fewer than two unmarked. The longest remaining low-velocity run is labelled fixation if it still reaches 20~ms; the rest of the original run is labelled as excluded for position range (subtype 15).

\subsection{Labelling, accounting, and self-checking}

Every detection stage assigns classes through a single function (P07) that writes only to samples not yet classified, so the order of stages fixes the precedence among them and no stage can overwrite another's decision. MNH, by contrast, sets its index vectors directly and mutably. Twenty-two subtype codes distinguish tracker loss from filter loss, pre- from post-blink exclusion, and each rejection test from every other (Table~\ref{tab:subtypes}). Candidate saccades and PSOs are written to per-recording decision files with the decision and its reason; PSO candidates shorter than 7~ms, PSOs dropped after logging, and saccade candidates at the very end of a recording are not logged. A consistency check (P24) verifies that each class has as many samples as its index vector, that PSOs do not outnumber saccades, that per-event records are aligned, and that no retained PSO falls below the floor. FKM also writes a main-sequence fit per recording and a saccade file for oculomotor-plant estimation \cite{komogortsev2014}.

\section{Parameters and output codes}\label{sec:params}

\begin{table}[!htbp]
\centering
\caption{Parameters of MNH and FKM}\label{tab:params}
\small
\begin{tabular}{@{}>{\raggedright\arraybackslash}p{0.40\textwidth}ll@{}}
\toprule
Parameter & MNH & FKM \\
\midrule
Savitzky--Golay order; window & 2; 7~ms & 2; 7~ms \\
Filter application & causal (\texttt{filter}) & centred convolution, edge-padded \\
Saccade peak-velocity threshold & 55~\degs & 55~\degs \\
Saccade on/off threshold & 45~\degs & 40~\degs \\
Step forward from local minimum at onset & up to 4 samples$^{a}$ & 3 samples \\
Step back from local minimum at offset (no PSO) & 3 samples & 3 samples \\
Minimum samples above peak threshold & $> 2$ & $> 2$ \\
Minimum saccade duration & 10~ms & 8~ms \\
Maximum duration of the peak block & --- & 105~ms \\
Small/large saccade amplitude criterion & --- & 8.08\textdegree \\
Artifact velocity ceiling & 1000~\degs & 680~\degs \\
Artifact acceleration ceiling & 100,000~\degss & 225,000~\degss \\
RIONEPS inefficiency threshold & 100 & not implemented \\
Provisional-fixation velocity ceiling for noise estimate & 100~\degs & 100~\degs \\
Minimum run and end trim for noise estimate & 30~ms; 5~ms & 20~ms; 3 samples \\
Fixation-noise threshold & 90th percentile & 90th percentile \\
Minimum fixation duration & 30~ms & 20~ms \\
Maximum position range within a fixation & 2.0\textdegree & 2.5\textdegree\ (salvaged if exceeded) \\
Minimum PSO duration & 3~ms & 7~ms \\
PSO velocity floor; grading thresholds & 20; 20 / 45 / 55~\degs & 40; 40 / 51 / 72~\degs \\
Maximum excess of PSO duration over saccade duration & --- & 17~ms \\
Maximum local velocity maxima in a PSO & --- & 6 \\
Noisy-recording flag (proportion of samples lost)$^{b}$ & 0.18 & 0.22 \\
\bottomrule
\end{tabular}
\tablenote{$^{a}$The archived MNH classifications use $\min(4, c - m)$ samples, where $c$ is the 45~\degs\ crossing and $m$ the local minimum; the public MNH code uses a fixed 4 (Section~\ref{sec:mnh}). $^{b}$Written to the output; does not change the classification.}
\end{table}

\begin{table}[!htbp]
\centering
\caption{FKM classes and subtype codes. Codes marked $^{\dagger}$ are defined in the code but never written}\label{tab:subtypes}
\small
\begin{tabular}{@{}>{\raggedright\arraybackslash}p{0.22\textwidth}
                  >{\raggedright\arraybackslash}p{\dimexpr0.78\textwidth-2\tabcolsep\relax}@{}}
\toprule
Class & Subtypes \\
\midrule
1 Fixation & --- \\
2 Saccade & 0; 4 large saccade (amplitude $>$ 8.08\textdegree) \\
3 PSO & 1 small, 2 moderate, 3 large \\
4 Noise from signal & 5 tracker NaN $\cdot$ 6 filter NaN $\cdot$ 9 before NaN block $\cdot$ 10 after NaN block $\cdot$ 13 out of screen range$^{\dagger}$ \\
5 Noise from detection & 0 unclassified residue $\cdot$ 7 velocity $>$ 680~\degs\ $\cdot$ 8 acceleration $>$ 225,000~\degss\ $\cdot$ 11 PSO not preceded by saccade $\cdot$ 12 unclassified$^{\dagger}$ $\cdot$ 14 fixation too short $\cdot$ 15 fixation position range too large $\cdot$ 16 blink without NaN $\cdot$ 17 PSO peaks not declining $\cdot$ 18 saccade velocity peak at first or last sample $\cdot$ 19 pre-saccade baseline too noisy $\cdot$ 20 NaN in saccade peak block $\cdot$ 21 saccade peak block too short $\cdot$ 22 saccade ends too late in recording$^{\dagger}$ \\
\bottomrule
\end{tabular}
\end{table}

\section{Reconstruction, verification, and corrections}\label{sec:recon}

\subsection{Running FKM}

FKM was run through a driver that sets the parameters of Table~\ref{tab:params}, initialises the data structure as the distribution does, and calls the algorithm functions P04, P06--P08, P10--P16, and P19--P21 without modification, on the five-digit positions stored with the April 2019 MNH classifications. Classifications were produced in MATLAB R2024a; the driver also runs in GNU Octave 8 with seven small compatibility functions. The driver was verified with the uncorrected code: the raw position stored in the distribution's example output (a 101,157-sample random-saccade recording classified by the author in MATLAB) was fed through the driver, and the author's classification was reproduced on every sample, with identical event counts and noise threshold. Each recording was processed at full length and scored on its first 26,000 samples.

\subsection{Corrections to the code}

Review of the code during the reconstruction found defects, all corrected before the evaluation and listed line by line in the distribution (FKM\_corrections.diff). Two changed classifications. In the fixation-salvage procedure the code first labelled the entire over-range run as excluded and then attempted to relabel the salvaged portion as fixation; because labels are written once, the relabelling did nothing, every over-range fixation was rejected in full, and a phantom entry was added to the fixation list (D12). Separately, the statement meant to select the longest low-velocity run selected the first (D1). Two could change classifications in principle but did not on these recordings: two of four branches of the slope-test endpoint snap tested the wrong sign (D5), and the post-hoc PSO sweep could not retract a label (D11). The rest are bookkeeping: the blink test on the peak block passed uninitialised indices (D6); the PSO grading thresholds were overwritten by the consistency check (D3); one noise count was doubled (D2); the large-saccade subtype was never written (D13); velocity artifacts were written as subtype 8 instead of 7 (D15); and a tied position extremum wrote a vector to the decision log (D14).

On the twenty recordings the corrections changed 0.5\% of samples; on the random-saccade example recording, where fixations exceed 2.5\textdegree\ far more often, 2.8\%. Against all four raters, the corrected FKM raised fixation F1 in 17 of 17 studies and saccade $\kappa$ in 16 of 20, each by less than 0.005 at the median; no other metric and no boundary displacement moved. The conclusions hold for the uncorrected code.

Several further defects are documented but not corrected, because correcting them would change the method being described: an accepted saccade adjacent to a noise sample is meant to be rejected, but the relabelling cannot overwrite it and the accompanying count correction runs only when diagnostic output is enabled (so enabling verbose output changes the classification); the overlap tests in P12 are skipped for the first accepted saccade and PSO; P21 leaves the per-event record of a PSO it removes; and P24 compares counts, not sample membership.

\subsection{Decision accounting}

Over the twenty full-length recordings FKM examined 5704 peak blocks and accepted 3496 saccades. Of the 2208 rejected, 1021 contained NaN, 442 lay inside an already accepted PSO, 302 were too brief, 196 lay inside an accepted saccade, 117 assembled to a saccade shorter than 8~ms, 68 failed the noisy-baseline test, 42 the refractory test, 9 contained NaN after assembly, 8 had already been labelled PSO, 1 had its velocity peak at an edge, and 2 reached the end of the recording. Table~\ref{tab:psoaccount} gives the fate of PSO candidates.

\begin{table}[!htbp]
\centering
\caption{Fate of FKM's PSO candidates in the twenty full-length recordings. Candidates shorter than 7~ms are discarded without a log entry}\label{tab:psoaccount}
\small
\begin{tabular}{@{}lr@{}}
\toprule
Stage & PSOs \\
\midrule
Logged candidates & 2134 \\
\quad rejected: continuation of the saccade (slope test, passes 1 / 2) & $-271$ (10 / 261) \\
\quad rejected: more than six velocity maxima & $-28$ \\
\quad rejected: contains noise & $-22$ \\
\quad rejected: velocity peaks not declining & $-2$ \\
Passed the PSO detector & 1811 \\
\quad removed: saccade rejected afterwards & $-48$ \\
\quad removed: duration exceeds the saccade's by more than 17~ms & $-14$ \\
\quad removed: peak velocity at or below 40~\degs & $-963$ \\
\quad removed: not preceded by a saccade (P21) & 0 \\
Final PSOs & 786 \\
\bottomrule
\end{tabular}
\end{table}

\section{MNH and the delay-corrected re-run}\label{sec:mnh}

\subsection{MNH as used}

MNH is described in \cite{friedman2018}; its code was deposited in 2017 \cite{friedman2017mnhcode}. The classifications analysed are those of an MNH run made in April 2019, stored with the raw positions. Three details of the code differ from the 2018 description: the artifact velocity ceiling is 1000~\degs\ (the paper gives 1500), a PSO ends after six rather than five quiet samples, and the public code steps a fixed four samples forward from the local minimum at saccade onset, whereas the paper describes a step that varies with the interval between the threshold crossing and the minimum.

We re-ran the public code on all twenty recordings at full length. As deposited, it reproduced the archived classifications on 98.86\% of scored samples. Nearly every saccade offset matched, but about 40\% of archived onsets lay one to four samples earlier. With a single line changed so that the onset step is $\min(4, c - m)$---the onset is placed four samples after the minimum or at the 45~\degs\ crossing, whichever is earlier, as the 2018 paper describes---the re-run reproduced the archived classifications on 99.06\% of samples, with medians within 0.003 of the archived values on every measure and median boundary displacements within 1~ms of the archived ones. Most of the remaining differences lie in artifact regions: the archived files use two RIONEPS codes that the public code does not write. We conclude that the April 2019 run used the variable onset step, and we treat that as MNH as used.

\subsection{Delay-corrected MNH}

The delay-corrected MNH is this re-run (public code with the variable onset step) with one further change in \texttt{calVelAcc\_sgolay.m}: the same filter coefficients are applied with the same \texttt{filter} call to a copy of the signal padded with $(F-1)/2 = 3$ replicates of its first and last values, and the output is realigned by $2 \times 3$ samples, so that each sample is estimated from the window centred on it. Signs and scaling are MNH's. No other line differs.

Moving the causal re-run's labels three samples earlier gives the same classification on 99.77\% of samples, and on every agreement measure the paired differences between the two constructions are exactly zero (Table~\ref{tab:mnhc}). The archived labels moved three samples earlier agree with the centred re-run on 98.97\% of samples. Subtracting the filter's group delay from MNH's boundaries is therefore equivalent to re-running it with a centred filter.

\begin{table}[!htbp]
\centering
\caption{Delay-corrected MNH: centred re-run, label shift, and comparisons with FKM. Recording-level median difference, 95\% bootstrap CI, recordings higher, and exact sign-test $p$; trained raters; common snippets}\label{tab:mnhc}
\small
\begin{tabular}{@{}>{\raggedright\arraybackslash}p{0.28\textwidth}lcccc@{}}
\toprule
Comparison & Measure & Median & 95\% CI & Higher & $p$ \\
\midrule
Centred re-run $-$ shifted causal re-run & every measure & 0.000 & [0.000, 0.000] & 0/0 & --- \\[3pt]
\midrule
Centred re-run $-$ causal re-run & Fixation $\kappa$ & $+0.029$ & $[+0.021, +0.037]$ & 20/20 & $<.001$ \\
 & Saccade $\kappa$ & $+0.029$ & $[+0.015, +0.045]$ & 19/20 & $<.001$ \\
 & PSO $\kappa$ & $+0.127$ & $[+0.117, +0.152]$ & 16/16 & $<.001$ \\
 & PSO F1 & $+0.247$ & $[+0.135, +0.308]$ & 16/16 & $<.001$ \\[3pt]
\midrule
FKM $-$ centred MNH & Fixation $\kappa$ & $-0.026$ & $[-0.033, -0.015]$ & 2/20 & $<.001$ \\
 & Fixation F1 & $+0.000$ & $[+0.000, +0.000]$ & 4/7 & 1.00 \\
 & Saccade $\kappa$ & $+0.000$ & $[-0.007, +0.013]$ & 10/20 & 1.00 \\
 & Saccade F1 & $+0.006$ & $[+0.000, +0.011]$ & 13/17 & .049 \\
 & PSO $\kappa$ & $-0.139$ & $[-0.278, -0.064]$ & 2/16 & .004 \\
 & PSO F1 & $-0.152$ & $[-0.296, -0.061]$ & 2/16 & .004 \\[3pt]
\midrule
FKM (20~\degs\ floor) $-$ centred MNH & Fixation $\kappa$ & $-0.008$ & $[-0.013, -0.002]$ & 4/20 & .012 \\
 & Fixation F1 & $+0.000$ & $[+0.000, +0.000]$ & 4/4 & .125 \\
 & Saccade $\kappa$ & $+0.000$ & $[-0.007, +0.013]$ & 10/20 & 1.00 \\
 & Saccade F1 & $+0.006$ & $[+0.000, +0.011]$ & 13/17 & .049 \\
 & PSO $\kappa$ & $-0.011$ & $[-0.031, +0.039]$ & 5/16 & .21 \\
 & PSO F1 & $-0.002$ & $[-0.036, +0.063]$ & 7/16 & .80 \\
\bottomrule
\end{tabular}
\tablenote{Pooled PSO precision / recall on the samples both classified: centred MNH 0.49 / 0.70; FKM 0.72 / 0.40; FKM with the 20~\degs\ floor 0.61 / 0.61. Boundary timing of centred MNH (saccade onset, saccade offset, PSO onset, PSO offset): $-1$, 0, 0, $+1$~ms.}
\end{table}



\section{Additional analyses}\label{sec:additional}

\subsection{Reference labels}

Raters' labels were mapped to the scored classes as follows: fixation (the default), saccade (regular and wavy), PSE, and other (blink, RIONEPS, unclassifiable); MNH's and FKM's noise classes map to other. Four people classified the recordings independently. The archived files identify raters only by number; we take rater 4 to be the training leader (the first author) because only raters 1--3 reproduce the three-rater agreement that \cite{friedman2023} reported without him (median $\kappa$ within 0.009 of that report, against 0.033 to 0.048 for any set of three including rater 4). The position files were matched to the anonymised subject numbers by comparing each rater's saccade onsets with MNH's: all twenty matched a unique recording with 83--100\% of onsets within 8~ms, against at most 20\% for any other. The rating tool read MNH's output files and is not archived; we cannot establish whether raters saw MNH's classification. Had they anchored on it, rater boundaries would have moved towards MNH's, reducing FKM's measured timing advantage.

\subsection{Agreement per rater, and with the designer}

\begin{table}[!htbp]
\centering
\caption{Median agreement of MNH (as used) and FKM with each rater}\label{tab:perrater}
\small
\begin{tabular*}{\textwidth}{@{\extracolsep{\fill}}lllccccc@{}}
\toprule
Class & Metric & Algorithm & Rater 1 & Rater 2 & Rater 3 & Trained & Rater 4 \\
\midrule
Fixation & $\kappa$ & MNH & 0.867 & 0.834 & 0.877 & 0.862 & 0.862 \\
 & & FKM & 0.877 & 0.857 & 0.869 & 0.867 & 0.871 \\
 & F1 & MNH & 0.982 & 0.978 & 0.984 & 0.982 & 0.983 \\
 & & FKM & 0.990 & 0.984 & 0.990 & 0.987 & 0.987 \\
 \midrule

Saccade & $\kappa$ & MNH & 0.876 & 0.861 & 0.873 & 0.870 & 0.883 \\
 & & FKM & 0.907 & 0.890 & 0.901 & 0.902 & 0.906 \\
 & F1 & MNH & 0.968 & 0.957 & 0.971 & 0.970 & 0.969 \\
 & & FKM & 0.964 & 0.953 & 0.965 & 0.963 & 0.963 \\
\midrule

PSO & $\kappa$ & MNH & 0.482 & 0.484 & 0.436 & 0.476 & 0.430 \\
 & & FKM & 0.469 & 0.483 & 0.484 & 0.482 & 0.435 \\
 & F1 & MNH & 0.255 & 0.408 & 0.236 & 0.278 & 0.298 \\
 & & FKM & 0.421 & 0.439 & 0.337 & 0.426 & 0.326 \\
\bottomrule
\end{tabular*}
\end{table}

FKM's saccade-$\kappa$ advantage over MNH was the same against each trained rater (median $+0.034$, $+0.034$, $+0.032$) and slightly smaller against rater 4 ($+0.029$). Rater 4 agreed with the trained raters as closely as they agreed with one another on fixations and saccades ($\kappa$ 0.946 and 0.967) and less closely on PSEs (0.756, against 0.803 between trained raters).

\subsection{Boundary timing in detail}

\begin{table}[!htbp]
\centering
\caption{Algorithm boundary minus rater boundary (ms), trained raters pooled, events matched at intersection-over-union $>$ 0.5}\label{tab:timing}
\small
\begin{tabular*}{\textwidth}{@{\extracolsep{\fill}}lcccc@{}}
\toprule
 & Matched & MNH median & FKM median & Both matched: \\
Boundary & (MNH / FKM) & [IQR] & [IQR] & MNH / FKM ($n$) \\
\midrule
Saccade onset & 4713 / 4861 & $+2$ [$+1$, $+3$] & $-1$ [$-3$, 0] & $+2$ / $-1$ (4651) \\
Saccade offset & 4713 / 4861 & $+2$ [0, $+4$] & 0 [$-2$, $+1$] & $+3$ / 0 (4651) \\
PSO onset & 910 / 862 & $+3$ [$+3$, $+4$] & 0 [0, $+1$] & $+3$ / 0 (599) \\
PSO offset & 910 / 862 & $+4$ [$+3$, $+5$] & $+2$ [$+1$, $+3$] & $+4$ / $+2$ (599) \\
\bottomrule
\end{tabular*}
\end{table}

\begin{figure}[!htbp]
\centering
\includegraphics[width=\textwidth]{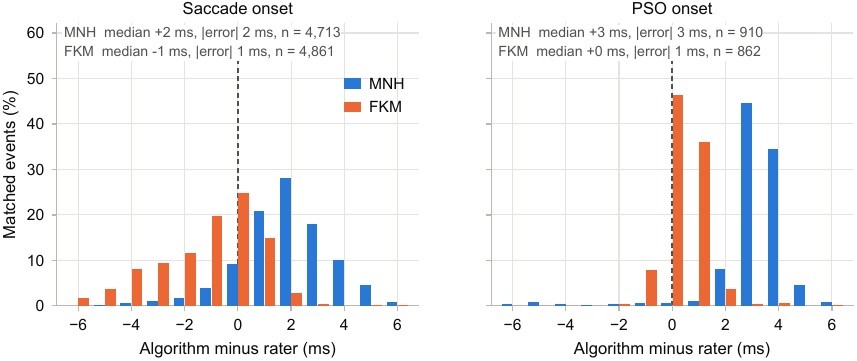}
\caption{Algorithm minus rater boundary at saccade onset (left) and PSO onset (right), events matched against the three trained raters. FKM places 46\% of PSO onsets on the rater's sample; MNH places 45\% three samples after it.}\label{fig:boundary}
\end{figure}

The median absolute error at least halved at every boundary. On saccades matched by both algorithms, the share of onsets within 1~ms of the rater's rose from 33.7\% to 59.3\%. The per-rater medians at saccade onset were the same for all four raters ($+2$~ms for MNH, $-1$~ms for FKM). FKM's onset distribution is skewed early: 17.9\% of onsets fall 3--4~ms before the rater's (Fig.~\ref{fig:boundary}).

\subsection{Wavy saccades: coverage}

Kept-whole wavy saccades did not always span the rater's interval: the saccade label covered a median 85.2\% of the interval under MNH and 94.4\% under FKM; the share kept whole with at least 80\% coverage was 50.7\% for MNH, 50.6\% for delay-corrected MNH, and 74.2\% for FKM. For wavy saccades kept whole, the saccade end was a median 2~ms late under MNH and 1~ms early under FKM. Counts are rater-event instances pooled over the three trained raters, so one physical saccade can be counted up to three times.

\subsection{Abstentions counted as disagreement}

When agreement was computed over every sample the rater classified, with an algorithm's unclassified samples counted as disagreement, the ordering was unchanged: median saccade $\kappa$ 0.868 (MNH), 0.886 (delay-corrected MNH, label shift), 0.898 (FKM); PSO $\kappa$ 0.415, 0.559, and 0.425.

\subsection{An external detector}

REMoDNaV \cite{dar2021}, version 1.1.2, was run at default settings on the same signals (fixation and pursuit mapped to fixation; all PSO labels to PSO). Its median saccade $\kappa$ (0.885) fell between MNH (0.870) and FKM (0.902); FKM's was higher on 14 of 20 recordings (median $+0.017$, 95\% CI $[-0.001, +0.040]$, $p = .12$). Its PSO F1 (0.456) was close to FKM's, with lower precision (0.43) and higher recall (0.46). Treating its pursuit labels (31\% of scored samples) as unclassifiable left the comparison unchanged.

\subsection{RIONEPS}\label{sec:rioneps}

The raters labelled 19 RIONEPS intervals totalling 1918 samples---0.10\% of the rated samples---in three recordings. Of those samples, MNH labelled 38\% noise and FKM 22\%; the absence of MNH's detector from FKM cost about 300 rater-labelled samples in 1,976,000 rater-sample observations.

\section{Code, data, and reproduction}\label{sec:repro}

All materials are available at \href{https://hdl.handle.net/10877/25355}{https://hdl.handle.net/10877/25355}. 
The package contains:
\begin{itemize}
\item \texttt{FKM\_original/}: FKM as left by its author (P01--P24, user guide).
\item \texttt{FKM\_paper\_supplement/FKM\_corrected/} and \texttt{FKM\_corrections.diff}: the corrected FKM and every changed line.
\item \texttt{FKM\_paper\_supplement/evaluation/}: the FKM driver (\texttt{run\_fkm\_headless.m}, \texttt{run\_fkm\_batch.m}), the ablation switches, the scorer (\texttt{score.py}, a line-by-line port of the original MATLAB scorer, checked against it on all 76 studies to $5\times10^{-5}$), the paired-test and bootstrap code, the REMoDNaV runner, and the saved classifications.
\item \texttt{mnh\_rerun/}: the public MNH code with the variable onset step (causal) and with the centred filter, the batch runner, the comparison script, and its output.
\item \texttt{GazeBaseR\_hand\_labels\_by\_rater/}: the hand classifications of raters 1--4, with loaders and the subject-to-recording mapping.
\item \texttt{Session TEX Class/}: the position signals and archived MNH classifications of the twenty recordings.
\end{itemize}

\textit{Software.} MATLAB R2024a with the Signal Processing Toolbox (FKM and MNH runs; FKM also runs in GNU Octave 8); Python 3.10 with pinned packages (\texttt{requirements.txt}) for all statistics and figures.

\textit{Reproduction.} \texttt{python reproduce\_all.py} regenerates every classification, number, and figure (about 15~min with MATLAB); \texttt{python reproduce\_all.py --python} re-scores the saved classifications only. The MNH re-run is reproduced with \texttt{run\_mnh\_rerun\_batch.m} and \texttt{mnh\_rerun\_check.py}.

\bibliographystyle{unsrt}
\bibliography{ms}